\UseRawInputEncoding

\documentclass[aps,prb,twocolumn,groupedaddress,showpacs, citeautoscript,superscriptaddress]{revtex4-1}
\usepackage{graphicx}
\usepackage{dcolumn}
\usepackage{bm}
\usepackage{amsmath,amsfonts,amssymb,amsthm,verbatim}
\usepackage[ansinew]{inputenc}
\usepackage{color}

\usepackage{epsfig}
\usepackage{wasysym}

\begin{document}

\title{Vacancy-Engineered Flat-Band Superconductivity in Holey Graphene}





\author{Matheus S. M. de Sousa}

\affiliation{Department of Physics, PUC-Rio, 22451-900 Rio de Janeiro, Brazil}

\author{Fujun Liu}

\affiliation{Instituto de F\'{i}sica, Universidade de Bras\'{i}lia, Bras\'{i}lia-DF, Brazil}

\author{Fanyao Qu}

\affiliation{Instituto de F\'{i}sica, Universidade de Bras\'{i}lia, Bras\'{i}lia-DF, Brazil}

\author{Wei Chen}

\affiliation{Department of Physics, PUC-Rio, 22451-900 Rio de Janeiro, Brazil}

\date{\today}

\begin{abstract}

A bipartite lattice with chiral symmetry is known to host zero energy flat bands if the numbers of the two sublattices are different. We demonstrate that this mechanism of producing flat bands can be realized on graphene by introducing periodic vacancies. Using first-principle calculations, we elaborate that even though the pristine graphene does not exactly preserve chiral symmetry, this mechanism applied to holey graphene still produces single or multiple bands as narrow as $\sim 0.5$eV near the Fermi surface throughout the entire Brillouin zone. Moreover, this mechanism can combine with vacancy-engineered nonsymmorphic symmetry to produce band structures with coexisting flat bands and nodal lines. A weak coupling mean-field treatment suggests the stabilization of superconductivity by these vacancy-engineered narrow bands. In addition, superconductivity occurs predominantly on the majority sublattices, with an amplitude that increases with the number of narrow bands.

\end{abstract}

\maketitle


\section{Introduction}

The superconductivity (SC) discovered recently in twisted bilayer graphene (TBLG) revises the interest on the search for SC on graphene based materials\cite{Cao18_2}. While the origin of the observed SC is still under intense debate, the small flat band due to splitting and anticrossing of the Dirac cones from the two graphene sheets is generally considered to play an important role\cite{Suarez10,Bistritzer11,Moon12,Laissardiere12,dosSantos12,Fang16,Cao18}. In fact, the possibility of flat band induced SC has long been of great interest, since the CuO$_{2}$ plane of high temperature superconductor materials\cite{Bednorz86,Lee06,Keimer15}, at which the SC occurs\cite{Anderson87,Emery87,Zhang88}, has the same structure (if Cu and O are not distinguished) as the Lieb lattice that is known to host a flat band and has been realized in various other systems\cite{Shen10,GuzmanSilva14,Mukherjee15,Taie15,Vicencio15,Julku16,Xia16,Diebel16,Slot17,Klembt17,Cui20}. On the other hand, from the density of states (DOS) point of view, it is intriguing to ask whether there exists some generic mechanisms that can generate flat bands in a large area of the Brillouin zone (BZ) of single-layer graphene at low energy, such that SC may be stabilized.





In this paper, we elaborate that the a generic mechanism of producing zero-energy flat bands (ZEFBs) on any bipartite lattice, first put forward by Lieb\cite{Lieb89}, can be realized on graphene by introducing periodic vacancies. Lieb's theorem states that on a bipartite lattice with chiral (sublattice) symmetry, ZEFBs occur if the numbers of the two sublattices per unit cell are different $N_{A}\neq N_{B}$, and the ZEFBs are at least $|N_{A}-N_{B}|$-fold degenerate. We demonstrate that this situation can be created on the honeycomb lattice by removing different numbers of the two sublattices in an enlarged unit cell, and ZEFBs throughout the entire BZ occur provided the tight-binding model of the lattice preserves the chiral symmetry. This mechanism is then tested in a realistic single-layer graphene by first-principle calculations. Our results indicate that despite graphene in reality does not exactly preserve chiral symmetry, this mechanism can still produce bands as narrow as $\sim 0.5$eV near the Fermi surface throughout the entire BZ, and moreover can be used to engineer an exotic band structure that contains both flat bands and nodal lines.

Our theoretical investigation is largely motivated by the fact that graphene with vacancies, often called holey graphene or graphene nanomesh, has been realized by various experimental techniques, such as nitrogenation\cite{Pawlak20,Mahmood15}, self-aligned anisotropic etching\cite{Shi11}, nano-network masking\cite{Jung14}, and lithography using copolymer\cite{Bai10}, nanosphere\cite{Wang13_2}, or He ion beam\cite{Archanjo14}, suggesting the feasibility of vacancy engineering in reality. Note that our proposal considers complete removal of carbon atoms, which is in contrast to the flat bands produced by the periodic potentials from adatoms\cite{Skurativska21}, requires much less removal of atoms compared to the cyclicgraphyne\cite{You19} or azite\cite{Shima93} proposals, and may also be realizable in superlattices nanolithographed in semiconductor thin films\cite{Tadjine16}. In addition, since these vacancy-engineered narrow bands dramatically enlarge the DOS at the Fermi surface compared to the pristine graphene, we examine the possibility of phonon-mediated SC by means of a weak coupling mean-field theory using the realistic phonon band width\cite{Saito01,Maultzsch04,Mohr07,Jorio11}. The results indeed point to the occurrence of SC in these vacancy configurations, whose spatial pattern is highly influenced by the chiral symmetric wave function of the ZEFBs, with a magnitude that increases with the number of ZEFBs.

The structure of the paper is organized in the following manner. In Sec.~II, we first revisit Lieb's theorem with an emphasis on the chiral symmetry of the ZEFB wave functions. Two vacancy configurations are then used to demonstrate perfect ZEFBs on a honeycomb lattice that preserves chiral symmetry. The narrow bands of these two configurations realized in graphene are then elaborated by first-principle calculations, and additionally another vacancy configuration that yields coexisting narrow bands and nodal lines. In Sec.~III, we lay out the weak-coupling mean field theory to investigate SC on the first two vacancy configurations, especially to detail their spatial pattern and dependence on the pairing interaction and degeneracy of the ZEFBs. These results are summarized in Sec.~IV.

\section{Vacancy-engineered flat bands on graphene} 

\subsection{Chiral symmetry and rank-nullity theorem \label{sec:chiral_rank_nullity}}

We first revisit Lieb's theorem that is based on the rank-nullity theorem\cite{Lieb89}, with a special emphasis on the nonspatial symmetries, localization of the wave functions, and applications to periodic vacancies. We consider any two- (2D) or three-dimensional (3D) bipartite-lattices described by single-particle Hamiltonian $H({\bf k})$ in momentum space that preserves time-reversal (TR), particle-hole (PH) chiral symmetries
\begin{eqnarray}
&&TH({\bf k})T^{-1}=H(-{\bf k}).
\nonumber \\
&&CH({\bf k})C^{-1}=-H(-{\bf k}).
\nonumber \\
&&SH({\bf k})S^{-1}=-H({\bf k})
\label{TCS_symmetries}
\end{eqnarray}
which are nonspatial symmetries particularly relevant to topological order\cite{Schnyder08,Ryu10,Chiu16,vonGersdorff21_unification} and topological phase transitions\cite{Chen19_universality_class}.
In these bipartite lattices, the Hamiltonian matrix arranged in the basis of the electron operators of the two sublattices $(c_{B{\bf k}},c_{A{\bf k}})$ is a block-off-diagonal $2\times 2$ matrix, and the symmetry operators are implemented by $T=K$, $C=\sigma_{3}K$ and $S=\sigma_{3}$, where $K$ is the complex conjugation operator.

\begin{figure}[ht]
\begin{center}
\includegraphics[clip=true,width=0.8\columnwidth]{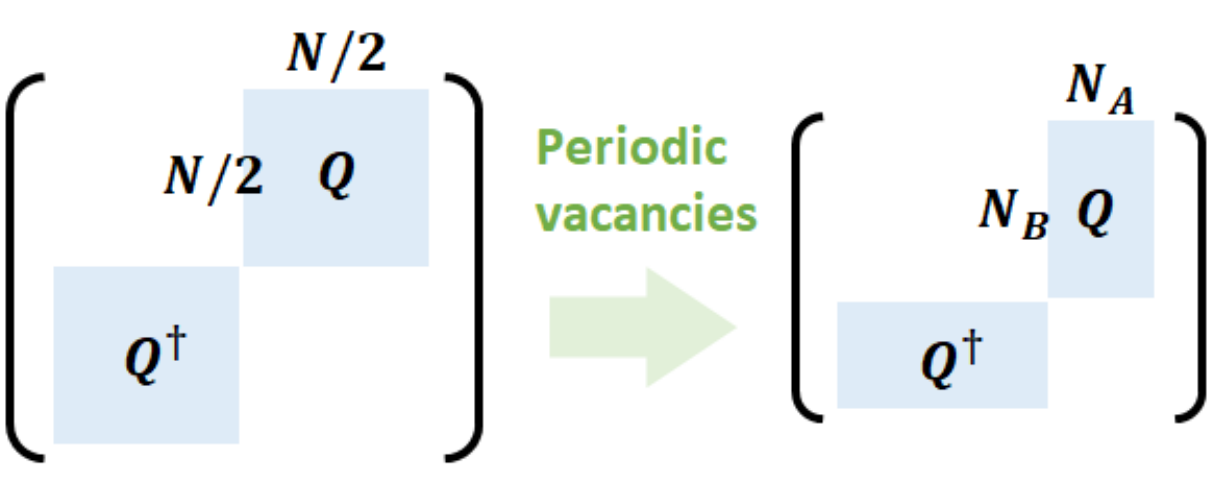}
\caption{Schematics of the proposed generic mechanism for ZEFBs. Starting from a chiral symmetric bipartite lattice, the Hamiltonian matrix $H({\bf k})$ describing an enlarged unit cell containing $N_{A}=N_{B}=N/2$ sites of each sublattice is block-off-diagonal (blue blocks). After periodic vacancies are introduced, which correspond to removing some columns and rows in the Hamiltonian, the off-diagonal block becomes not square if the unit cell contains different numbers of the two sublattices $N_{B}>N_{A}$. In this case, the rank-nullity theorem ensures that ZEFBs must occur. } 
\label{fig:Hk_not_square_schematics}
\end{center}
\end{figure}

Now suppose we enlarge the unit cell to contain not 2 but $N=$ even number of sites with the same amount of two sublattices, then the Hamiltonian matrix arranged in the basis $(c_{B_{1}{\bf k}},...,c_{B_{N/2}{\bf k}},c_{A_{1}{\bf k}},...,c_{A_{N/2}{\bf k}})$ remains block-off-diagonal 
\begin{eqnarray}
H({\bf k})=\left(\begin{array}{cc}
 & Q({\bf k}) \\
Q^{\dag}({\bf k}) & \\
\end{array}\right),
\label{Hk_honeycomb}
\end{eqnarray}
where $Q({\bf k})$ is an $(N/2)\times (N/2)$ square matrix. The symmetry operators are implemented by $C=\sigma_{3}\otimes I_{N/2}K$ and $S=\sigma_{3}\otimes I_{N/2}$ in this case, with $I_{N/2}$ the $(N/2)\times (N/2)$ identity matrix. If we now introduce periodic vacancies into the lattice, the columns and rows in the $H({\bf k})$ in Eq.~(\ref{Hk_honeycomb}) that correspond to the vacancy sites will be removed. It is then obvious that if the number of vacancies on the $A$ and $B$ sublattices are different, then the unit cell will contain different number of sublattices $N_{A}\neq N_{B}$. As a result, the $Q({\bf k})$ in Eq.~(\ref{Hk_honeycomb}) as an $N_{A}\times N_{B}$ matrix is not a square matrix anymore, as elaborated in Fig.~\ref{fig:Hk_not_square_schematics}. Nevertheless, the PH and chiral symmetries of the systems still hold since removing the columns and rows in $C=\sigma_{3}\otimes I_{N/2}K$ and $S=\sigma_{3}\otimes I_{N/2}$ that correspond to the vacancy sites preserve Eq.~(\ref{TCS_symmetries}). As a result, the band structure at any vacancy configuration is PH symmetric, and the ZEFB wave functions must be localized on one of the two sublattices since they are eigenstates of the chiral operator $S$. In fact, we will prove below that the ZEFB wave functions must localize in the majority sublattices.


Denoting $r(M)$ as the rank and $\eta(M)$ as the nullity of a matrix $M$, our interest is how the nullity of the Hamiltonian $\eta(H)$, which counts the number of ZEFBs, can be nonzero. Without loss of generality, we assume that the vacancy configuration is such that the $B$ sublattices are the majority $N_{B}>N_{A}$. In this case, the rank-nullity theorem states that
\begin{eqnarray}
r(Q)+\eta(Q)=N_{A},
\;\;\;r(H)+\eta(H)=N_{A}+N_{B}.
\label{rank_nullity_H_Q}
\end{eqnarray}
For $H$ of the form of Eq.~(\ref{Hk_honeycomb}), the rank satisfies
\begin{eqnarray}
&&r(Q)=r(Q^{\dag})=r(QQ^{\dag})=r(Q^{\dag}Q),
\nonumber \\
&&r(H)=r(Q)+r(Q^{\dag})=2r(Q).
\label{rank_identities}
\end{eqnarray}
Using these simple identities in linear algebra, we now prove the following propositions. 

{\it Proposition 1: $\eta(H)>0$ if $Q$ is not square.} This can be proved easily from Eqs.~(\ref{rank_nullity_H_Q}) and (\ref{rank_identities})
\begin{eqnarray}
\eta(H)=N_{A}+N_{B}-2r(Q)=N_{B}-N_{A}+2\eta(Q).
\end{eqnarray}
Hence $\eta(H)>0$ if $N_{B}>N_{A}$, meaning that ZEFB must emerge if we remove the two sublattices in different quantities.

{\it Proposition 2: $\eta(H)=N_{B}-N_{A}$ if $\eta(Q)=0$.} To prove this, we start from 
\begin{eqnarray}
r(QQ^{\dag})+\eta(QQ^{\dag})=N_{B},\;\;\;
r(Q^{\dag}Q)+\eta(Q^{\dag}Q)=N_{A},\;\;\;\;\;
\label{rank_nullity_QQdag}
\end{eqnarray}
which implies 
\begin{eqnarray}
r(QQ^{\dag})-r(Q^{\dag}Q)=N_{B}-N_{A}.
\end{eqnarray}
If $Q$ itself is not singular $\eta(Q)=0$, which is true in many practical examples, then Eq.~(\ref{rank_nullity_H_Q}) implies $r(Q)=N_{A}=r(QQ^{\dag})=r(Q^{\dag}Q)$. Then from Eq.~(\ref{rank_nullity_QQdag}) one sees that $\eta(QQ^{\dag})=N_{B}-N_{A}$ and $\eta(Q^{\dag}Q)=0$. Because the square of the Hamiltonian $H^{2}={\rm diag}(QQ^{\dag},Q^{\dag}Q)$ has the same nullity and ZEFBs wave functions as $H$, we immediately see that $\eta(H)=\eta(H^{2})=\eta(QQ^{\dag})=N_{B}-N_{A}$. The propositions 1 and 2 constitute the original version of Lieb's theorem.

{\it Proposition 3: ZEFB wave functions are localized in the majority sublattices if $\eta(Q)=0$.} This is simply because the proof for proposition 2 shows that the ZEFB wave functions are given by diagonalizing the first block $QQ^{\dag}$ in $H^{2}$ that belongs to the majority $B$ sublattices, whereas wave functions in $A$ sublattices are zero.

\begin{figure}[ht]
\begin{center}
\includegraphics[clip=true,width=0.99\columnwidth]{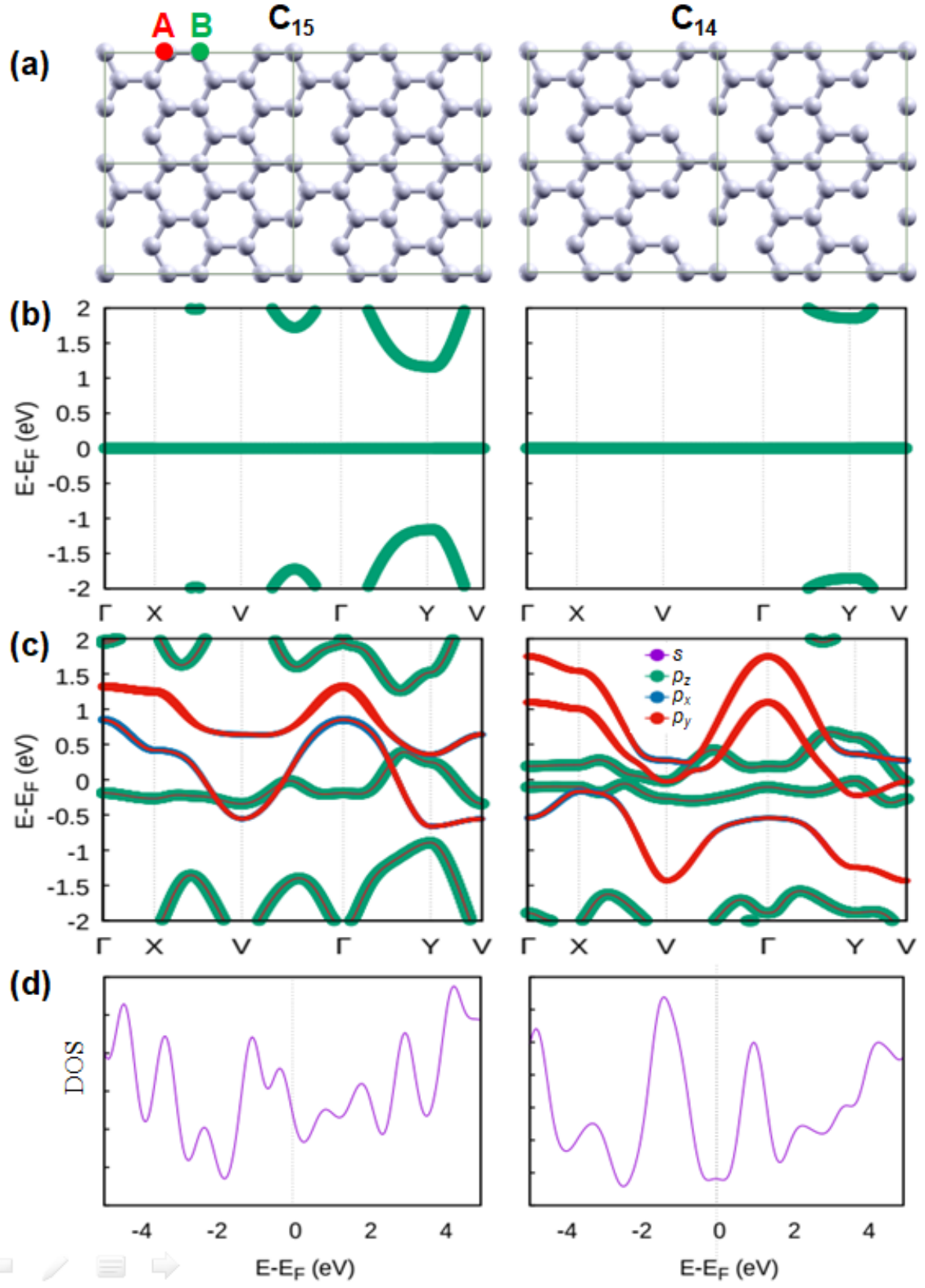}
\caption{Numerical results for the two vacancy configurations C$_{15}$ (left column) and C$_{14}$ (right column), whose lattice structures are shown in (a)  with the two sublattices $A$ and $B$ indicated. (b) The band structures obtained from the nearest-neighbor tight-binding model that preserves chiral symmetry, which contains a single ZEFB for C$_{15}$ and doubly degenerate ZEFBs for C$_{14}$. (c) DFT band structures and (d) the corresponding DOS for these two vacancy configurations realized by graphene that does not exactly preserve the chiral symmetry, which still yield very narrow bands of mainly $p_{z}$ at low energy. } 
\label{fig:tight_binding_example}
\end{center}
\end{figure}




We aim to examine these propositions in the spinless honeycomb lattice with nearest-neighbor hopping 
\begin{eqnarray}
H=\sum_{\langle ij\rangle}t\,c_{i}^{\dag}c_{j}+U\sum_{i\in v}c_{i}^{\dag}c_{i}
\end{eqnarray}
owing to its relevance to the $p_{z}$ orbital of single-layer graphene that will be addressed later, where $c_{i}$ denotes the electron annihilation operator at site $i$, and $U$ is the large on-site potential that can be used to conveniently project out the vacancy sites $i\in v$. Increasing from $U=0$ to $U>100t$ can continuously change the band structure from Dirac points to ZEFBs, similar to that proposed recently in the Lieb-kagome lattices\cite{Lim20}, but we will focus on the large on-site potential regime $U>100t$ that completely removes the vacancy sites. We denote these vacancy configurations by C$_{N_{A}+N_{B}}$. In the left panel of Fig. \ref{fig:tight_binding_example} (a), we show a C$_{15}$ example of removing a single $A$ sublattice from a $N=16$ rectangular unit cell such that $N_{A}=7$ and $N_{B}=8$. In this case, the BZ is rectangular, and the PH symmetric tight-binding band structure plotted along a high-symmetry line (HSL) clearly indicates a single ZEFB throughout the BZ, as shown in the left panel of Fig.~\ref{fig:tight_binding_example} (b). In contrast, the C$_{14}$ configuration shown in the right panel of Fig.~\ref{fig:tight_binding_example} (a) and (b) that removes two $A$ sublattices on the same 16-site unit cell, such that $N_{A}=6$ and $N_{B}=8$, has doubly degenerate ZEFBs $\eta(H)=N_{B}-N_{A}=2$, consistent with proposition 2. In Appendix \ref{apx:ZEFB_wave_fn}, we also elaborate proposition 3 by presenting the wave functions of the ZEFBs for C$_{15}$ and C$_{14}$ and show that indeed both are localized on the majority $B$ sublattices. Finally, we remark that although we focus on periodic vacancies on an infinite graphene, the propositions 1 to 3 also explain the number of zero eigenenergies and their wave functions of a finite size graphene with random vacancies\cite{Bouzerar21}.

\subsection{Application to realistic graphene} 

To examine the proposed mechanism on the realistic single-layer graphene, it should be first noted that graphene in reality does not exactly preserve the PH and chiral symmetries in Eq.~(\ref{TCS_symmetries}) owing to the complications such as longer range hopping and hybridization between different orbitals, although magnitudes of these factors are much smaller than the nearest-neighbor hopping\cite{CastroNeto09}. To investigate the effect of these symmetry breaking factors, we turn to density functional theory (DFT) to obtain the band structure of graphene with periodic vacancies. In Fig.~\ref{fig:tight_binding_example} (c), we show the DFT band structures and DOS for C$_{15}$ and C$_{14}$, which indicate that very narrow bands do occur near the Fermi surface. Although not perfectly flat, these bands are as narrow as $\sim 0.5$eV, reminisce the ZEFBs. Moreover, the DOS at the Fermi surface is dramatically enhanced by these narrow bands compared to the pristine graphene, as indicated by Fig.~\ref{fig:tight_binding_example} (d). 


\subsection{Coexistence of ZEFBs and nodal lines}

\begin{figure}[ht]
\begin{center}
\includegraphics[clip=true,width=0.99\columnwidth]{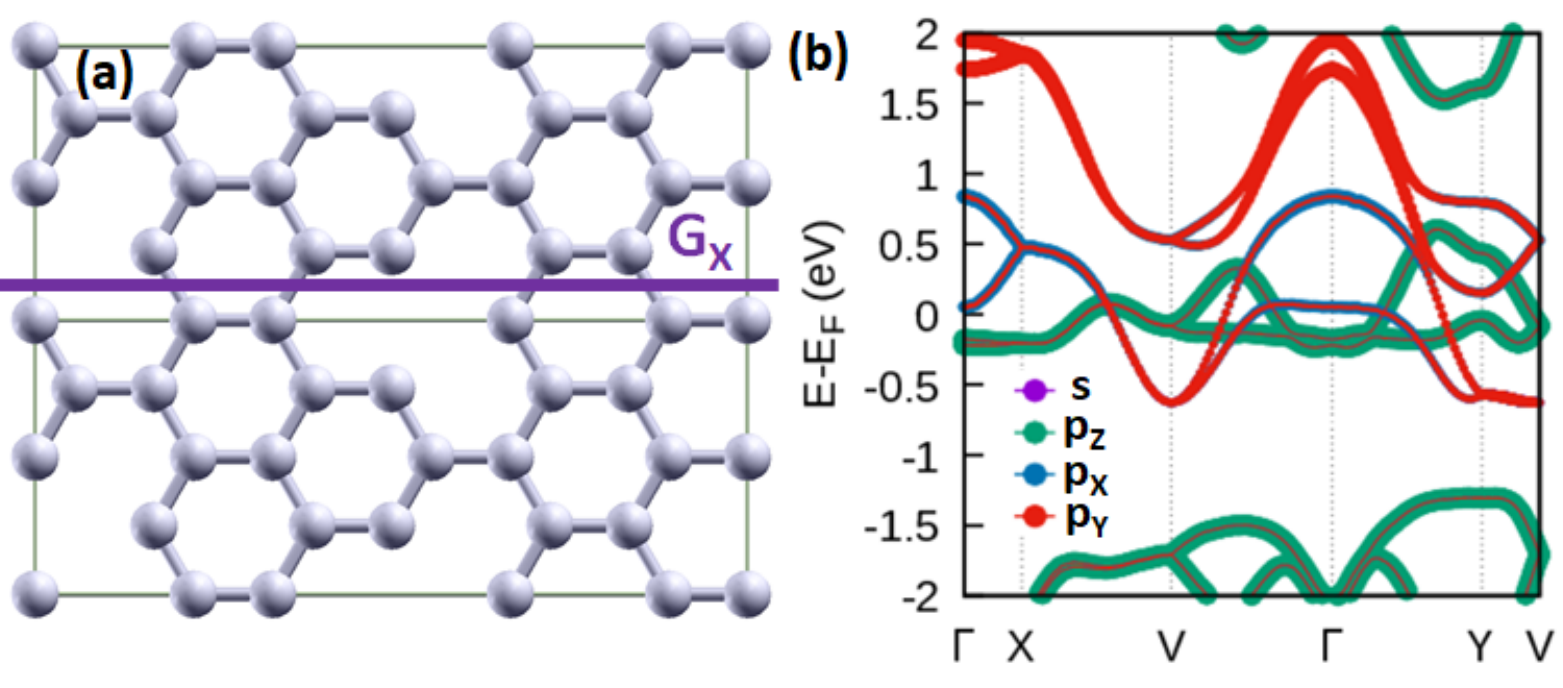}
\caption{(a) A vacancy configuration C$_{22}$ that contains two missing $A$ sublattices and a glide plane $G_{x}$. (b) The resulting band structure contains two low energy narrow bands of mainly $p_{z}$ orbital origin, and in addition every two pairs of spin degenerate bands stick together to form four-fold degenerate nodal lines at the $X-V$ section of the BZ boundary. } 
\label{fig:nodal_lines_ZEFB}
\end{center}
\end{figure}

We proceed to demonstrate that the proposed mechanism can coexist with another vacancy engineering principle proposed recently, namely the nodal-line semimetals caused by 2D nonsymmorphic vacancy configurations\cite{Liu21,Liu21_2,deSousa21_nodal_line}. In these nonsymmorphic configurations, every two carbon atoms map to each other under glide plane operation, resulting in nodal lines at the BZ boundary regardless the details of the Hamiltonian, which serves as a vacancy-engineering principle to obtain symmetry-enforced nodal lines\cite{Young15,Yamakage16,Zhao16,Wieder18}. As an example, in Fig.~\ref{fig:nodal_lines_ZEFB} we show a C$_{22}$ configuration that contains two vacancies on the $A$ sublattice, and additionally a glide plane along ${\hat{\bf x}}$ direction. The resulting DFT band structure indicates that both mechanisms prevail in this situation, yielding a band structure that contains two low energy narrow bands that reminisce doubly degenerate ZEFBs, and in addition every two pairs of bands (each pair is spin degenerate) stick together at the BZ boundary $k_{x}=0$ (the $X-V$ section of Fig.~\ref{fig:nodal_lines_ZEFB} (b)) to form symmetry-enforced nodal lines as predicted. This example indicates that vacancy engineering can combine different crystalline and nonspatial symmetries to produce very exotic band structures that may not be easily found in nature.

\section{Mean-field treatment of SC stabilized by ZEFBs}

\subsection{Bogoliubov-de Gennes equations for graphene with periodic vacancy} 

The enlarged DOS near the Fermi surface due to the narrow bands is expected to dramatically change the electronic and magnetic properties of the holey graphene compared to the pristine graphene. However, these narrow bands are neither perfectly flat nor entirely located at zero energy, and hence it is unclear at present whether strong correlations will play an important role on the electronic properties in the same way as in TBLG\cite{Cao18_2}. On the other hand, the phonon band width in graphene is about $\omega_{D}\approx 0.25$eV\cite{Saito01,Maultzsch04,Mohr07,Jorio11}, which means that the narrow bands with band width $\sim 0.5$eV in a large area of the BZ are within the Debye frequency, suggesting a large phase space for phonon-mediated Cooper pairing\cite{Kopnin11,Heikkila16}. These features motivate us to examine whether a conventional phonon-mediated $s$-wave SC phase emerges in the holey graphene with low energy narrow bands. For this purpose, we resort to the following spinful mean-field model of $s$-wave SC
\begin{eqnarray}
H&=&\sum_{\langle ij\rangle\sigma}t\,c_{i\sigma}^{\dag}c_{j\sigma}+\sum_{\langle\langle ij\rangle\rangle\sigma}t'\,c_{i\sigma}^{\dag}c_{j\sigma}-\sum_{i\sigma}\mu\,c_{i\sigma}^{\dag}c_{i\sigma}
\nonumber \\
&&+\sum_{i}\left(\Delta_{i}c_{i\uparrow}^{\dag}c_{i\downarrow}^{\dag}
+\Delta_{i}^{\ast}c_{i\downarrow}c_{i\uparrow}\right)+U\sum_{i\in v}c_{i\sigma}^{\dag}c_{i\sigma},
\label{mean_field_SC_Hamiltonian}
\end{eqnarray}
where $t=2.8$eV is the nearest-neighbor hopping on the honeycomb lattice, and the on-site potential $U>100 t$ is used to project out the vacancy sites $i\in v$. We use the next-nearest-neighbor hopping $t'$ and chemical potential $\mu$ to simulate the breaking of chiral symmetry in realistic graphene, and find that the values $t'=-0.2$eV and $\mu=0.2$eV can give a reasonable fit to the narrow bands obtained by DFT in both C$_{15}$ and C$_{14}$, as demonstrated in the Appendix \ref{apx:ZEFB_wave_fn}. The mean field Hamiltonian is diagonalized into $H={\rm const.}+\sum_{\bf k\alpha}E_{\bf k}\gamma_{\bf k\alpha}^{\dag}\gamma_{\bf k\alpha}$ by a Bogoliubov transformation
\begin{eqnarray}
&&c_{i\uparrow}=\sum_{\bf k}\gamma_{\bf k\uparrow}u_{\bf k}(i)-\gamma_{\bf k\downarrow}^{\dag}v_{\bf k}^{\ast}(i),
\nonumber \\
&&c_{i\downarrow}=\sum_{\bf k}\gamma_{\bf k\downarrow}u_{\bf k}(i)+\gamma_{\bf k\uparrow}^{\dag}v_{\bf k}^{\ast}(i),
\end{eqnarray}
where $i=1,2...N_{A}+N_{B}$ denotes the site inside a unit cell, and $\gamma_{\bf k\sigma}$ is the annihilation operator of the Bogoliubov quasiparticles. The wave functions $\left\{u_{\bf k}(i),v_{\bf k}(i)\right\}$ and eigenenergy $E_{\bf k}$ satisfy
\begin{eqnarray}
E_{\bf k}u_{\bf k}(i)&=&\sum_{\langle ij\rangle}t\,u_{\bf k}(j)+\sum_{\langle\langle ij\rangle\rangle}t'\,u_{\bf k}(j)
\nonumber \\
&&+U\delta_{i\in v}u_{\bf k}(i)+\Delta_{i}v_{\bf k}(i),
\nonumber \\
E_{\bf k}v_{\bf k}(i)&=&-\sum_{\langle ij\rangle}t\,v_{\bf k}(j)-\sum_{\langle\langle ij\rangle\rangle}t'\,v_{\bf k}(j)
\nonumber \\
&&-U\delta_{i\in v}v_{\bf k}(i)+\Delta_{i}^{\ast}u_{\bf k}(i).
\label{BdG_u_v_equations}
\end{eqnarray}
After the Hamiltonian is diagonalized, the pairing amplitude at site $i$ is calculated by
\begin{eqnarray}
\Delta_{i}=\sum_{\bf k}V\theta(\omega_{D}-E_{\bf k})\left[2f(E_{\bf k})-1\right]u_{\bf k}(i)v_{\bf k}^{\ast}(i),
\label{BdG_gap_eq}
\end{eqnarray}
where $f(E_{\bf k})=(e^{E_{\bf k}/k_{B}T}+1)^{-1}$ is the Fermi distribution and $V<0$ is the pairing interaction acting within Debye frequency $\omega_{D}=0.25$eV, as ensured by the step function $\theta(\omega_{D}-E_{\bf k})$. Equations (\ref{BdG_u_v_equations}) and (\ref{BdG_gap_eq}) are solved self-consistently until the local pairing amplitude $\Delta_{i}$ converges at a given pairing interaction and temperature $\left\{V,T\right\}$. 


\begin{figure}[ht]
\begin{center}
\includegraphics[clip=true,width=0.99\columnwidth]{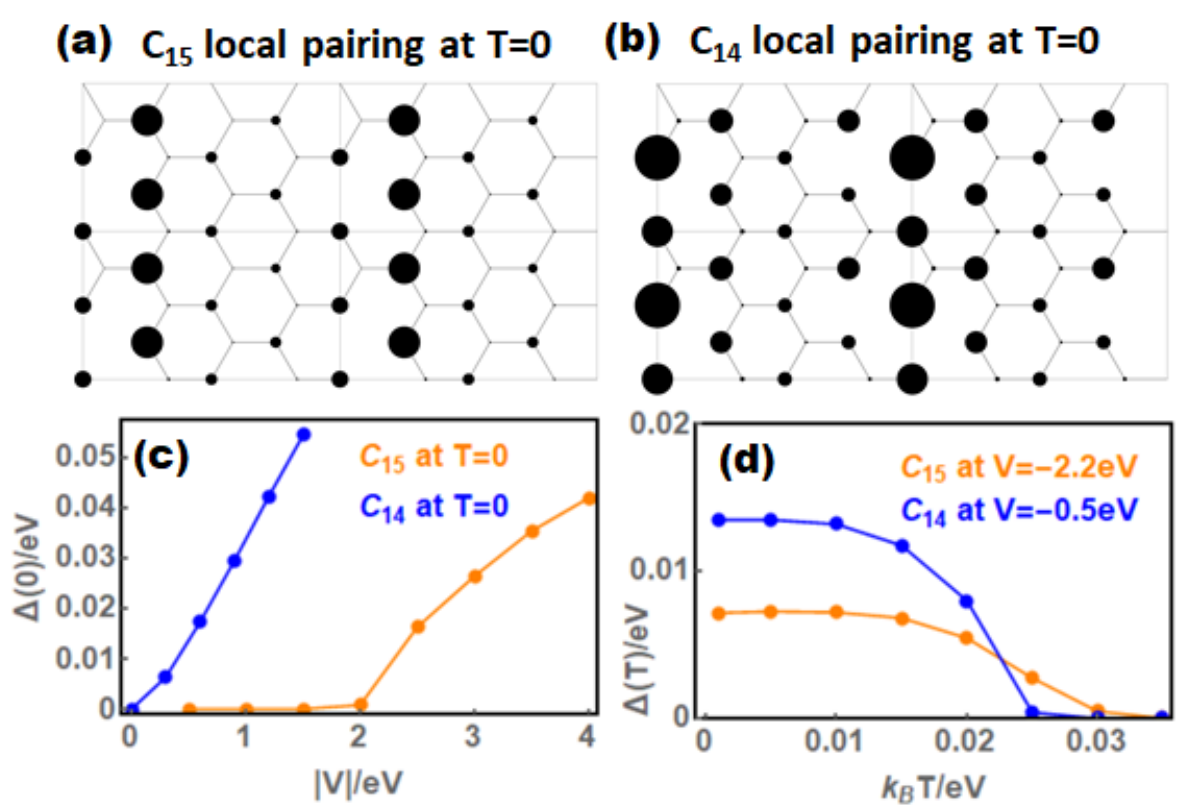}
\caption{(a) Local pairing amplitude $\Delta_{i}$ represented by circle size for the C$_{15}$ configuration in Fig.~\ref{fig:tight_binding_example}, calculated at zero temperature and pairing interaction $V=-2.2$eV by our weak coupling mean-field model. The largest circles correspond to $\Delta_{i}=0.036$eV. (b) $\Delta_{i}$ for the C$_{14}$ configuration in Fig.~\ref{fig:tight_binding_example} at zero temperature and pairing interaction $V=-0.5$eV, where the largest circles correspond to $\Delta_{i}=0.074$eV. In these two figures, one sees that the $B$ sublattices that the vacancy sites do not belong to have much larger pairing amplitude. (c) Spatially averaged pairing amplitude at zero temperature $\Delta(0)$ versus the pairing interaction $|V|$. (d) The spatially average pairing amplitude $\Delta(T)$ as a function of temperature. } 
\label{fig:SC_C15_C14}
\end{center}
\end{figure}

\subsection{Numerical results for the local pairing}

Numerical calculation of our mean-field theory applied to the pristine graphene yields $\Delta_{i}=0$ everywhere, consistent with the experimental observation that the single-layer pristine graphene has no SC. In contrast, Fig.~\ref{fig:SC_C15_C14} (a) and (b) show the local gap $\Delta_{i}$ at zero temperature for the two vacancy configurations C$_{15}$ and C$_{14}$ illustrated in Fig.~\ref{fig:tight_binding_example}. Our result indicates a finite $\Delta_{i}$ in the holey graphene, and moreover $\Delta_{i}$ on the $B$ sublattices is about two orders of magnitude larger than that on the $A$ sublattices. This feature stems from the fact that the chiral symmetry breaking terms are relatively small $\left\{t',\mu\right\}\ll t$, so the narrow band wave functions still roughly satisfy the proposition 3 in Sec.~\ref{sec:chiral_rank_nullity} and hence are mainly localized on the $B$ sublattices, as discussed in Appendix \ref{apx:ZEFB_wave_fn}. For C$_{15}$, the spatially averaged gap $\Delta(T)=\sum_{i}\Delta_{i}/(N_{A}+N_{B})$ at zero temperature $\Delta(0)$ is vanishingly small at small pairing potential $V$. Only when the pairing potential has the same order of magnitude as the hopping $|V|\sim t$ does a sizable gap emerge, implying that a sufficiently strong electron-phonon interaction is needed to support SC in C$_{15}$. On the other hand, C$_{14}$ requires much smaller $|V|$ to trigger SC in comparison with C$_{15}$, suggesting that increasing the number of narrow bands does help to create the SC phase. Concerning the temperature dependence, the spatially averaged gap shows a trend similar to the usual weak coupling superconductors, which vanishes at a critical temperature $T_{c}$ that is higher at larger pairing potential $V$. In addition, $T_{c}$ is generally raised in the configurations with more narrow bands, consistent with that expected from an enlarged DOS. Since the proposed engineering mechanism in principle has no restriction on the number of narrow bands $N_{B}-N_{A}$ (times $2$ if including spin), we anticipate that the vacancy configurations with very different numbers of the two sublattices may yield a very high $T_{c}$, which is presumably more likely to engineer in configurations with a larger unit cell.


\section{Conclusions}

In summary, we demonstrate that Lieb's theorem of ZEFBs can be realized on graphene by introducing periodic vacancies, which can be experimentally relevant to holey graphene or graphene nanomesh. Although graphene in reality does not preserve the chiral symmetry required by the theorem, periodic vacancies can still induce bands as narrow as $\sim 0.5$eV near the Fermi surface because the symmetry breaking factors are relatively weak compared to the nearest-neighbor hopping. Moreover, our results suggest that periodic vacancies can be used to combine various nonspatial and crystalline symmetries to produce very exotic band structures, such as the coexisting ZEFBs and nodal lines revealed in the present work, paving a way to engineer band structures of 2D materials beyond the limitation set by the underlying crystalline structures.

The vacancy-engineered narrow bands dramatically enlarge the DOS near the Fermi surface, and hence are expected to significantly change the electronic and magnetic properties of the holey graphene compared to the pristine one, which await further investigations. In particular, our mean-field theory survey reveals that a phonon-mediated conventional SC is feasible due to the enlarged DOS. The local pairing amplitude is highly localized on the majority sublattices, a feature originated from the chiral symmetry of the flat band wave functions. The minimal electron-phonon interaction required to create the SC phase varies significantly with the number of narrow bands, but increasing the numbers of narrow bands generally reduces the minimal electron-phonon interaction, as expected from the DOS point of view. As a result, we anticipate that a certain experimental effort to search for the appropriate vacancy configuration that has a sufficient number of narrow bands is needed to observe the SC.


\appendix

\section{Flat band wave functions with and without chiral symmetry \label{apx:ZEFB_wave_fn}}

\begin{figure}[ht]
\begin{center}
\includegraphics[clip=true,width=0.99\columnwidth]{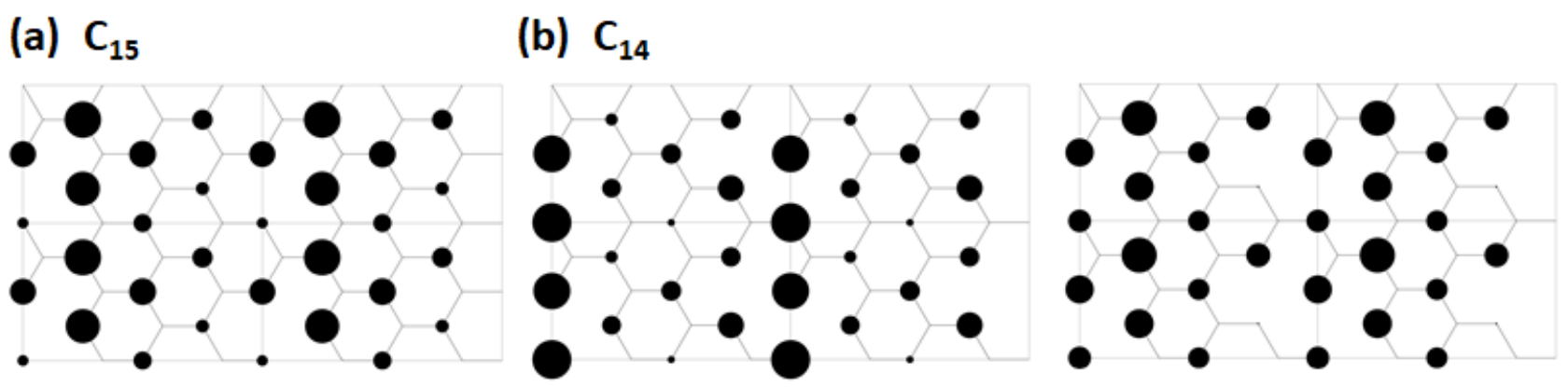}
\caption{Wave functions $|\psi_{i}|^{2}$ at momentum ${\bf k}=(0.15,0.37)$ of (a) the single ZEFB of C$_{15}$ and (b) the two degenerate ZEFBs of C$_{14}$ with only nearest-neighbor hopping, whose band structure is that shown in Fig.~\ref{fig:tight_binding_example} (b). The largest circles correspond to $|\psi_{i}|^{2}=0.313$. All these wave functions preserve chiral cymmetry and hence are localized only on the majority $B$ sublattices. } 
\label{fig:chiral_wave_fn_C15_C14}
\end{center}
\end{figure}

The proposition 3 in Sec.~\ref{sec:chiral_rank_nullity} states that the ZEFB wave functions for a chiral symmetric system in any vacancy configuration must localize on the majority sublattices. As an example, in Fig.~\ref{fig:chiral_wave_fn_C15_C14} we show the single flat band wave function for C$_{15}$ and the two degenerate flat band wave functions for C$_{15}$ at momentum ${\bf k}=(0.15,0.37)$, both described by a spinless nearest-neighbor hopping Hamiltonian $H=\sum_{\langle ij\rangle}t\,c_{i}^{\dag}c_{j}$ that preserves chiral symmetry. We find that all these wave functions are localized in the majority $B$ sublattices where the vacancies do not belong to, satisfying the proposition 3 in Sec.~\ref{sec:chiral_rank_nullity}.

\begin{figure}[ht]
\begin{center}
\includegraphics[clip=true,width=0.99\columnwidth]{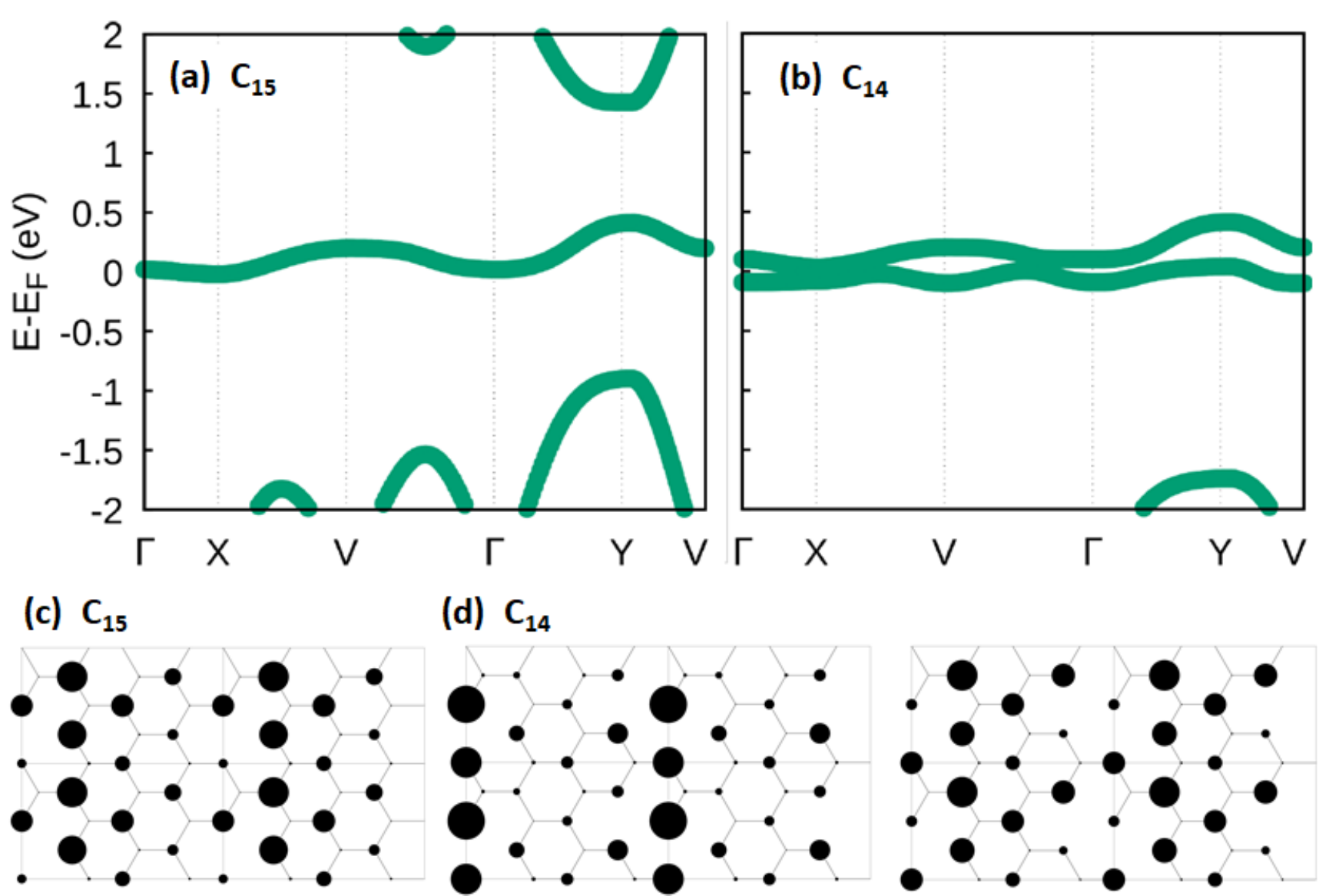}
\caption{Tight-binding band structure of (a) C$_{15}$ and (b) C$_{14}$ with $t=2.8$eV, $t'=-0.2$eV, and $\mu=0.2$eV, which yields a reasonable fit to the $p_{z}$ orbital bands obtained from DFT shown in Fig.~\ref{fig:tight_binding_example} (c). The wave functions $|\psi_{i}|^{2}$ of the narrow bands at momentum ${\bf k}=(0.15,0.37)$ are shown in (c) and (d), which are still highly localized on the $B$ sublattices since the chiral symmetry breaking terms are relatively weak $\left\{t',\mu\right\}\ll t$. } 
\label{fig:tight_binding_fit_C15_C14}
\end{center}
\end{figure}

For the realistic graphene that does not preserve chiral symmetry, we rely on numerical calculation to investigate the wave functions. Using a tight-binding model with nearest $t$ and next-nearest-neighbor hopping $t'$, and additionally a chemical potential $\mu$ (described by the Hamiltonian in Eq.~\ref{mean_field_SC_Hamiltonian} without pairing and spin), we find that the parameters $t=2.8$eV, $t'=-0.2$eV and $\mu=0.2$eV can fit the $p_{z}$ orbital bands of the DFT band structure shown in Fig.~\ref{fig:tight_binding_example} (c) reasonably well, as shown in Fig.~\ref{fig:tight_binding_fit_C15_C14} (a) for C$_{15}$ and (b) for C$_{14}$. The wave functions of the narrow bands close to Fermi surface are shown in Fig.~\ref{fig:tight_binding_fit_C15_C14} (c) and (d) at the same momentum ${\bf k}=(0.15,0.37)$ as that shown in Fig.~\ref{fig:chiral_wave_fn_C15_C14} (a) and (b). We find that because these chiral symmetry breaking terms $\left\{t',\mu\right\}$ are relatively small compared to $t$, the wave functions on the majority $B$ sublattices are about two orders of magnitude larger than that on the minority $A$ sublattices. In other words, the wave functions still roughly preserves the chiral symmetry and approximately satisfies the proposition 3 in Sec.~\ref{sec:chiral_rank_nullity}. As a result, the local pairing amplitude in the SC phase is much larger on $B$ sublattices, as shown in Fig.~\ref{fig:SC_C15_C14}.

\bibliography{Literatur}

\end{document}